\documentstyle[prl,aps,epsf]{revtex}
\begin{document}
\draft
\twocolumn[\hsize\textwidth\columnwidth\hsize\csname
@twocolumnfalse\endcsname
\title{Plastic flow, voltage bursts, and vortex avalanches
in superconductors}
\author{C. J. Olson$^{1}$, C. Reichhardt$^{1}$, J. Groth$^{1}$, 
Stuart B. Field$^{1,2}$, and Franco Nori$^{1,*}$}
\address{$^{1}$Department of Physics, The University of Michigan, Ann Arbor,
Michigan 48109-1120}
\address{$^{2}$Department of Physics, Colorado State University, Fort Collins,
Colorado 80523$^{\dagger}$}
\date{\today}
\maketitle
\begin{abstract}

We use large-scale parallel simulations to compute the motion of
superconducting magnetic vortices during
avalanches triggered by small field increases.
We find that experimentally observable 
voltage bursts correspond to pulsing vortex movement along branched channels
or winding chains, and relate vortex flow images to
features of statistical distributions. As pin density is increased, 
a crossover occurs from 
interstitial motion in narrow easy-flow winding channels
with typical avalanche sizes, to pin-to-pin motion in broad
channels, characterized by a very broad distribution of sizes.
Our results are consistent with recent experiments.
\end{abstract}
\vspace{-20pt}
\vskip2pc]
\vskip2pc
\narrowtext

Recent studies of systems characterized by avalanche dynamics, where
energy dissipation occurs in sudden bursts of collective activity,
have revealed many interesting dynamical behaviors.
Granular assemblies \cite{1}, 
magnetic domains \cite{2}, 
fluid flow \cite{3}, and flux lines
in type-II superconductors \cite{4,5,6,7,8}
all display avalanches, but produce broad distributions of
avalanche sizes only under certain conditions, indicating a lack
of universal response to small perturbations.
Probing the {\it microscopic} origins of different avalanche
behaviors by experimentally recording the microscopic motion
over time scales long enough to statistically characterize
the system is difficult. In contrast,
numerical simulations provide both precise control of microscopic
parameters and dynamical information, making them a useful tool.
Early avalanche simulations used simple discrete models
\cite{9}, but recent advances in parallel processing allow the use of more 
realistic continuous molecular dynamics (MD) models.
In this paper, we present extensive parallel MD simulations of 
flux-gradient-driven superconducting vortex avalanches.
By varying the pinning density $n_{p}$
and maximum pinning strength $f_{p}$, we obtain avalanche
distributions in good agreement with recent experiments
\cite{4,7}, and quantify how the density and strength of pinning 
sites affects the breadth of 
these distributions.  Characteristic avalanche sizes and lifetimes
appear at low pin densities when narrow easy-flow winding
channels of interstitial vortices form.
At higher pin densities, pin-to-pin transport through vortex
chains occurs, and the avalanche size and lifetime 
distributions remain very broad.  
We find that there is no universal distribution valid for all pinning 
parameters.

Flux penetrates a type-II superconductor in the form of discrete
quantized vortices.  As an external field
is slowly increased,
a metastable gradient in vortex density, termed the
Bean state \cite{10}, forms as vortices are driven into
the sample by their mutual repulsion and are held back
by defects in the material \cite{4,5,6,7,8}.
To model this system,
we simulate an infinite slab with a magnetic field ${\bf H}=H{\bf \hat{z}}$
applied parallel to the surface so that there are no demagnetization effects.
The rigid vortices and straight columnar pins we consider
are all parallel to ${\bf \hat{z}}$, so
we can obtain all relevant dynamical
information from a transverse two-dimensional
slice, in the $x$--$y$ plane, of the three-dimensional slab.
The flux lines evolve according to a $T=0$ MD algorithm.
The vortex-vortex repulsion, given by the modified Bessel
function $K_{1}(r/\lambda)$, is cut off
beyond $r=6\lambda$, where $\lambda$ is the penetration depth,
so that each vortex interacts with up to 100 neighbors, and important
{\it collective} effects, neglected in simulations with shorter interaction
ranges, are observed.  Each $24\lambda \times 26\lambda$
sample contains up to 3700 attractive parabolic
pins of radius $\xi_{p}=0.15\lambda$ with 
pinning densities $n_{p}=0.96/\lambda^{2}$, $n_{p}=2.40/\lambda^{2}$,
or $n_{p}=5.93/\lambda^{2}$,
and pinning strengths uniformly distributed over the range 
$f_{p}^{\rm max}/5$ to $f_{p}^{\rm max}$, where
$f_{p}^{\rm max}=0.3f_{0}$, $1.0f_{0}$, or $3.0f_{0}$.
All forces are given in units of
$f_{0}=\Phi_{0}^{2}/8\pi^{2}\lambda^{3}$ and lengths in units of $\lambda$.  

The total force on vortex $i$ is given by 
${\bf f}_{i}={\bf f}_{i}^{vv} + {\bf f}_{i}^{vp}=\eta{\bf v}_{i}$,
where the force on vortex $i$ from other vortices is 
${\bf f}_{i}^{vv}=\sum_{j=1}^{N_{v}} f_{0} K_{1} (|{\bf r}_{i}
-{\bf r}_{j} |/\lambda ) {\bf {\hat r}}_{ij}$, and
the force from pinning sites is
${\bf f}_{i}^{vp}=\sum_{k=1}^{N_{p}} (f_{p}/\xi_{p})
|{\bf r}_{i} - {\bf r}_{k}^{(p)}| \ \Theta ( (\xi_{p} - |{\bf r}_{i} - 
{\bf r}_{k}^{(p)} |) / \lambda ) {\bf {\hat r}}_{ik}.$
Here, $\Theta$ is the Heaviside step function,
${\bf r}_{i}$ (${\bf v}_{i}$) is the location (velocity) of vortex $i$,
${\bf r}_{k}^{(p)}$ is the location of pinning site $k$,
%$\xi_{p}$ is the pin radius, 
there are $N_{p}$ pinning sites and
$N_{v}$ vortices,
${\bf {\hat r}}_{ij}=({\bf r}_{i}-{\bf r}_{j})/
|{\bf r}_{i}-{\bf r}_{j}|$,
${\bf {\hat r}}_{ik}=({\bf r}_{i}-{\bf r}_{k}^{(p)})/
|{\bf r}_{i}-{\bf r}_{k}^{(p)}|$,
and we take $\eta=1$.
Using roughly $10^{4}$ hours on an IBM SP parallel computer,
we recorded more than $10^{4}$ avalanches for each of
five combinations of $n_{p}$ and $f_{p}$.
In addition to the results described in this paper,
we obtained distributions of
vortex velocities and displacements as well as
analytical results supporting our observations \cite{11}.
Further simulation details appear in \cite{11,12}. 

A slowly increasing external field is modeled
by adding a single vortex to an unpinned region
along the sample edge whenever the system reaches mechanical 
equilibrium \cite{11,12,13}.
The majority of these small field increases 

\begin{figure}
\centerline{
\epsfxsize=3.5 in
\epsfbox{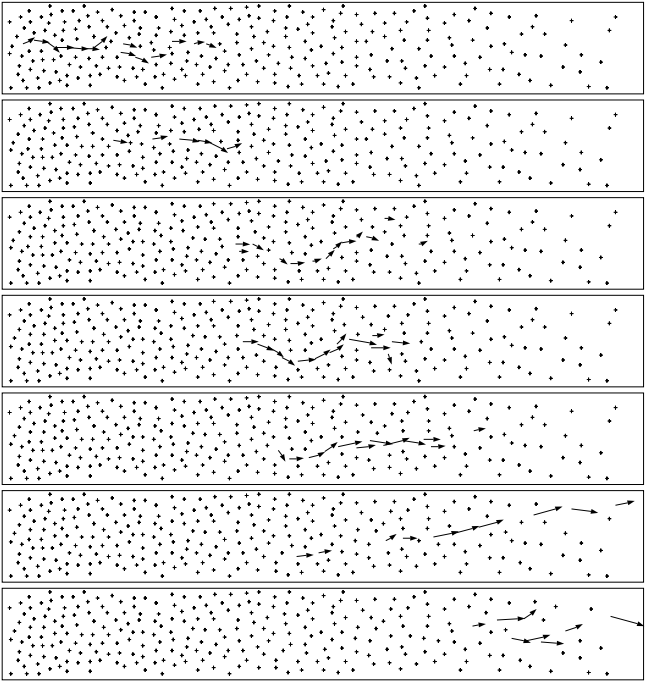}}
\caption{Consecutive snapshots of the velocity field of the vortex lattice
in a $25 \lambda \times
5 \lambda$ region of a $26\lambda \times 24\lambda$ 
sample with a high density of strong pinning
sites: $n_{p}=5.93/\lambda^{2}$, $f_{p}^{\rm max}=3.0f_{0}$.
Each moving vortex is indicated by an arrow with its length scaled by the
velocity of the vortex;  the remaining vortices are indicated by
small crosses.  All vortices sit in pinning sites (not shown here)
when not in motion.  The disturbance propagates from the dense left
edge of the sample to the relatively less dense right edge.  The
vortices in the rest of the sample (not shown) were not depinned.
The time interval between snapshots is the typical time $t_{h}$
required for a vortex to hop from one pinning site to another.  The
illustrated motion is typical for a medium-size avalanche in this sample.}
\label{fig:1}
\end{figure}

\hspace{-13pt}
result in only slight
shifts in vortex positions, but occasionally one or more vortices
will become depinned, producing an avalanche.
We find that avalanche disturbances propagate as an uneven pulse, as seen
in Fig.~\ref{fig:1} where arrows indicate instantaneous
vortex velocities.  Events
with longer lifetimes often contain more than one pulse of motion
(i.e., multiple oscillations in the total avalanche velocity)
\cite{14}.

\begin{figure}
\centerline{
\epsfxsize=3.5 in
\epsfbox{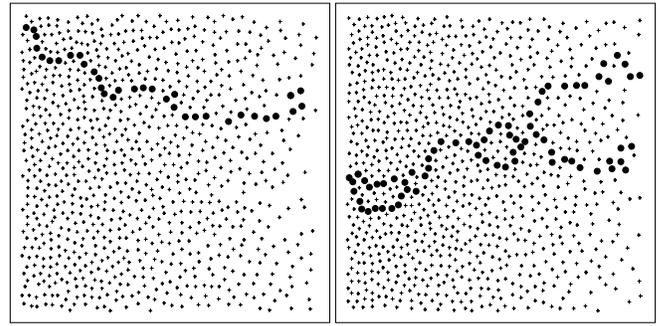}}
\caption{Snapshots of vortices depinned in an avalanche 
for two different events in the same sample.  
Here, stationary vortices are represented by crosses, while
the initial positions of vortices that were depinned
are marked with filled circles;  the pinning sites are not shown.
Vortices move towards the right of each panel, in the direction
of decreasing vortex density.  The entire
$26 \lambda \times 24 \lambda$ sample, with a pin density
of $n_{p}=2.40/\lambda^{2}$ and $f_{p}^{\rm max}=3.0f_{0}$, is shown.  
}
\label{fig:2}
\end{figure}

By imaging individual avalanches in our samples,
we find that a chain of vortices is displaced in a typical event.
Each vortex in the chain is depinned, moves a
short distance, and comes to rest in a nearby pinning site.  Vortices
outside the chain transmit stress by shifting very
slightly inside pinning sites, 
but are not depinned.  In Fig.~\ref{fig:2}, filled circles represent
the initial positions of vortices that were depinned
during the avalanche, while small crosses mark the
vortices that remained pinned. Easily visible chains of moving
vortices extend down the flux gradient in the $x$--direction, 
winding slightly in the $y$--direction.  Chain size varies from event 
to event:  in 

\hspace{-13pt}
some cases a chain spans the sample, as shown in 
Fig.~\ref{fig:2}, while in other events the chain
contains only three or four vortices.  
In each case, although the disturbance may cross the sample, an individual
vortex does not.  Thus, the time span of a typical avalanche is much
shorter than the time a single vortex takes to traverse the sample.

\begin{figure}
\centerline{
\epsfxsize=3.5 in
\epsfbox{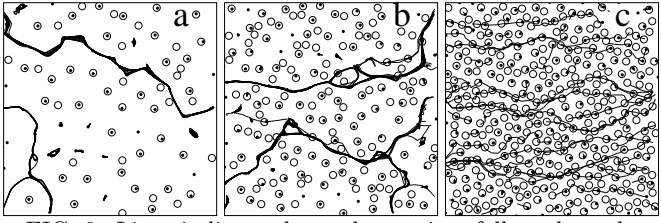}}
\caption{Lines indicate the paths vortices follow through a portion of the
sample over an extended period of time covering many avalanches. Here,
filled dots represent vortices, and open circles mark pinning sites.
Each panel shows an $8\lambda \times 8\lambda$ region of 
$24\lambda \times 26\lambda$ samples with 
$f_{p}^{\rm max}=3.0f_{0}$ and different pinning densities:
$n_{p}=0.96/\lambda^{2}$ (a),
$n_{p}=2.40/\lambda^{2}$ (b), and
$n_{p}=5.93/\lambda^{2}$ (c).
The presence or absence of easy-flow channels is clearly dependent on
pin density.  The channels present in (a) and (b) lead to avalanches with
characteristic sizes and lifetimes superimposed on
a broad distribution.  Samples with
higher pinning density produce very broad distributions of avalanche sizes.}
\label{fig:3}
\end{figure}

When the pinning density is high, chains of moving vortices are equally likely
to form anywhere in the sample.  As the pinning
density is lowered, vortices move only through narrow well-defined
interstitial winding channels in which 
vortices are weakly held in place only by the repulsion of other vortices
which sit in pinning sites.
To identify the cumulative pattern of flow channels for
different pinning parameters, in Fig.~\ref{fig:3} 
we plot vortex trajectories with lines over the course of many avalanches.
A concentration of trajectory lines indicates a heavily-travelled region 
of the sample.  In Fig.~\ref{fig:3}(a) we show a
small portion of a sample with low pinning density.
We see that all motion occurs through 
narrow {\it easy-flow interstitial channels}.  In these channels,
mobile interstitial vortices move {\it plastically} \cite{15}
around their strongly pinned neighbors 
in a manner similar to that recently imaged
experimentally \cite{6}.  This sample
also contains vortices oscillating in interstitial ``magnetic traps.''
As the pin density increases, the
amount of interstitial pinning decreases and the
number of channels increases, as in Fig.~\ref{fig:3}(b),
until at high pin densities [Fig.~\ref{fig:3}(c)], there is no interstitial
flow and the vortices move only from pin to pin. 
Here, where no easy-flow channel exists,
avalanches are spread evenly throughout the sample.

We determine how the microscopic 
pinning parameters affect avalanche size by finding
the total avalanche lifetime $\tau$ for each event.  Since vortices
typically hop from pin to pin in samples with high
pinning density, a natural unit of time is the interval $t_{h}$ a vortex spends
hopping between pinning sites, and so we use scaled lifetimes
$\tau^{*}=\tau/t_{h}$.  To find $t_{h}$, we assume that 

\begin{figure}
\centerline{
\epsfxsize=3.5 in
\epsfbox{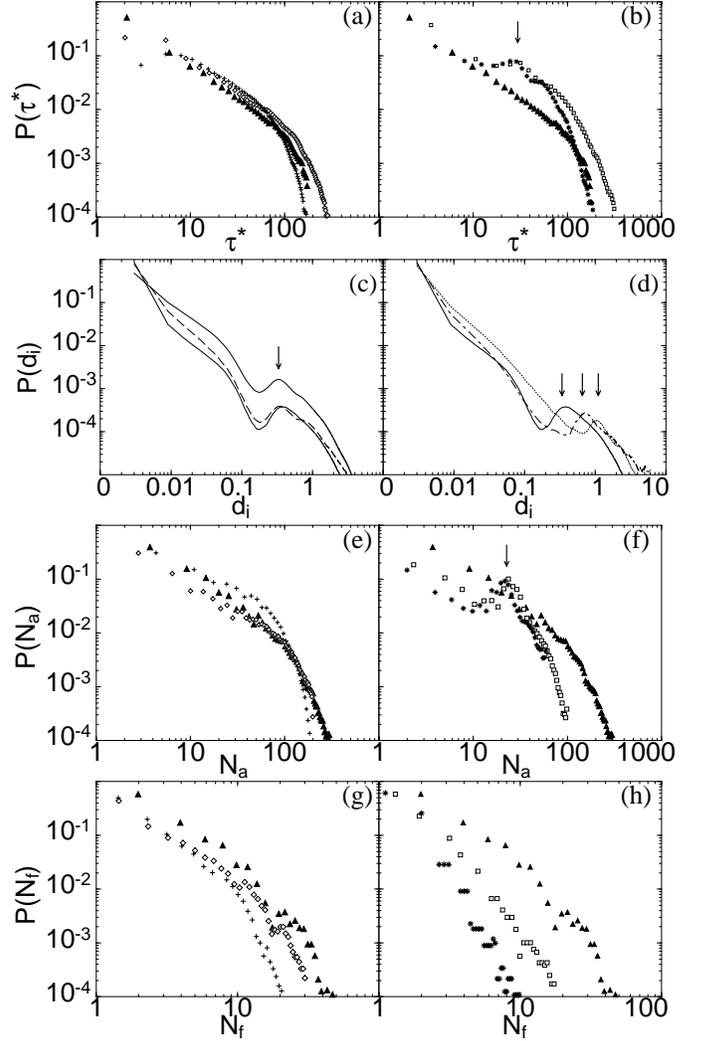}}
\caption{
Avalanche distributions:  lifetimes $\tau^{*}$ (a--b);
individual vortex displacements $d_{i}$ (c--d);
total number $N_{a}$ of vortices moving during an event (e--f);
number $N_{f}$ of vortices falling off the edge of the sample (g--h).
The left panels (a,c,e,g) refer to samples with high pin density
$n_{p}=5.93/\lambda^{2}$, and differing pinning strengths:  
filled triangles (solid line),  $f_{p}^{\rm max}=3.0f_{0}$;
open diamonds (dashed line),  $f_{p}^{\rm max}=1.0f_{0}$;
plus signs (heavy solid line), $f_{p}^{\rm max}=0.3f_{0}$.
The right panels 
(b,d,f,h) refer to samples with $f_{p}^{\rm max}=3.0f_{0}$  and
varying pinning densities:
filled triangles (solid line), $n_{p}=5.93/\lambda^{2}$;
open squares (dot-dashed line), $n_{p}=2.40/\lambda^{2}$;
asterisks (dotted line), $n_{p}=0.96/\lambda^{2}$.
In (a) the hopping time 
is $t_{h}=$ 6, 18, and 57 MD steps, respectively, while
in (b), $t_{h}=$ 6, 17, and 53 MD steps, respectively.}
\label{fig:4}
\end{figure}

\hspace{-13pt}
each vortex hops 
a distance $d_{p}= n_{p}^{-1/2}$, the average distance between 
pinning sites, and that
the vortex speed $v_{c}$ is proportional to the depinning force,
$v_{c}=|{\bf f}_{i}|/\eta $, where $|{\bf f}_{i}| \approx -f_{p}$.
This gives 
$t_{h}=d_{p}/v_{c} \approx \eta f_{p}^{-1} n_{p}^{-1/2}.$
In PbIn, for example, $\eta \sim 0.3$ G-m/$\Omega$, so using
$f_{p} = 3.0 f_{0}$, $n_{p} = 5.93/\lambda^{2}$, and
$\lambda \sim 65$ nm gives $t_{h} \sim 150$ ns.
The plot of $P(\tau^{*})$ in Fig.~\ref{fig:4}(a) for samples
with dense pinning shows that 
in each case the distribution is very broad and can be written 
as $P(\tau^{*}) \sim (\tau^{*})^{-1.4}$ over a range of $\tau^{*}$.  
If the pinning density $n_{p}$ 
is reduced, the form of $P(\tau^{*})$ changes noticeably, as
in Fig.~\ref{fig:4}(b).  Here, we find an enhanced probability
for avalanches with a characteristic value
%$\tau^{*} \approx T_{\rm char}$ (indicated by an arrow), 
as a result of the appearance of easy-flow interstitial channels.
Many avalanches in these samples consist of a single sample-spanning
pulse of motion through one of these channels.
The estimated pulse lifetime produced by a straight channel is
$\tau_{\rm est} \approx t_{h} N_{h}$,
where $N_{h}$ is the number of vortices in the channel.
Since $N_{h} \approx L_{x}\sqrt{n_{v}}$, where $L_{x}$
is the sample length, we find 
$\tau_{\rm est} \approx \eta L_{x} f_{p}^{-1}\sqrt{n_{v}/n_{p}}$, which
gives a characteristic value of
$\tau^{*} = \tau_{\rm est}/t_{h} \approx L_{x}\sqrt{n_{v}}=26\lambda \sqrt{1.5/\lambda^{2}}\approx 30$.
This value, indicated by an arrow in Fig.~\ref{fig:4}(b),
agrees well with the peak in the distribution of $\tau^{*}$.

For all pin densities, only a small
fraction of the vortices move significantly while the rest
shift in pinning sites, as seen by considering
the distance $d_{i}$ each vortex is displaced in an avalanche.
Vortices that hop from pin to pin create
a peak in $P(d_{i})$, marked with arrows in Fig.~\ref{fig:4}(c--d),
at $d_{i} \approx d_{p} = n_{p}^{-1/2}$.
For those vortices that remain pinned and accumulate
stress in the vortex lattice,
$d_{i}<d_{p}$, we can approximate the distribution 
by $P(d_{i}) \sim d_{i}^{-\rho}$, where $\rho \sim 1.4$ for 
all samples except $\rho \sim 1.2$ for
[$n_{p}=0.96/\lambda^{2}$, $f_{p}^{\rm max}=3.0f_{0}$], and
$\rho \sim 0.9$ for [$n_{p}=5.93/\lambda^{2}$,
$f_{p}^{\rm max}=0.3f_{0}$].  
An analytical argument \cite{11}, sketched here,
predicts a similar $\rho$ for all samples since 
the distribution is generated only by pinned 

\hspace{-13pt}
vortices and is not 
affected by the presence of easy-flow channels.
The addition of a vortex to the sample exerts a small additional force $f$ on
an arbitrary vortex located a small distance $r$ away $(r\ll \lambda)$, 
displacing this vortex a distance 
$\delta u(r)=(f/\eta) \delta t \sim K_{1}(r/\lambda) \sim 1/r$.  Since there
are $\delta N(r)=2\pi r n_{v} \delta r$ vortices a distance $r$ from the
added vortex,  we find 
$\rho=-d \ln \delta N / d \ln \delta u=-(1/r)(-1/r)^{-1}=1$,
in general agreement with our computed values of $\rho$,
$\rho \approx 0.9$ -- $1.4$.

Pinning density $n_{p}$ has a significant impact on 
the number of vortices $N_{a}$ that move
a distance $d_{i}$ larger than the pin diameter $2\xi_{p}$.
We plot $P(N_{a})$ in Fig.~\ref{fig:4}(e--f).
When $n_{p}$ is low, $n_{p} \lesssim 2.4/\lambda^{2}$, 
vortices move only in narrow easy-flow winding channels,  producing
a characteristic value of $N_{a}$, $N_{a} \approx N_{h}$, 
marked with an arrow in Fig.~\ref{fig:4}(f).  
The strongly pinned vortices around the interstitial channel cannot
be depinned, so large values of $N_{a}$ are not observed.
At high pinning densities, as in Fig.~\ref{fig:4}(e) where
$n_{p}=5.93/\lambda^{2}$,
we can approximate the form of $P(N_{a})$ for small $N_{a}$ 
as $P(N_{a}) \sim N_{a}^{-\beta}$.  
As the pin strength decreases, $\beta$ decreases slightly:
$\beta \sim 1.4$ for $f_{p}^{\rm max}=3.0f_{0}$,
$\beta \sim 1.0$ for $f_{p}^{\rm max}=1.0f_{0}$, and
$\beta \sim 0.9$ for $f_{p}^{\rm max}=0.3f_{0}$.
This is in contrast to the lifetime distributions which
were not affected by changing pinning strength.
Smaller values of $\beta$ indicate a relative increase in the frequency of
large avalanches compared to small ones.  The trend in $\beta$ 
indicates the importance of avalanche width
in determining $N_{a}$.  In samples with strong pinning, avalanche width
is suppressed since strongly pinned vortices on either side of the moving chain
are not depinned. As the pinning weakens,
wider avalanches occur when weakly pinned vortices adjacent to a slowly moving
chain depin and join the motion.  Thus, avalanches
with similar scaled lifetimes are likely to be wider in samples
with weak pinning than in samples with strong pinning.

Altering the pinning parameters affects
the number of vortices $N_{f}$ that exit the sample during an avalanche,
as shown in the plot of $P(N_{f})$ in
Fig.~\ref{fig:4}(g--h).  If we approximate $P(N_{f})$
for low $N_{f}$ by the form $P(N_{f}) \sim N_{f}^{-\alpha}$, we find that
all samples with high pinning density, $n_{p}=5.93/\lambda^{2}$, 
have $\alpha \sim 2.4$.  As $n_{p}$ decreases, $\alpha$
increases: $\alpha \sim 3.4$ for $n_{p}=2.40/\lambda^{2}$ and
$\alpha \sim 4.4$ for $n_{p}=0.96/\lambda^{2}$.  When
all vortex motion occurs in an easy-flow interstitial 
channel, $\alpha$ increases since the channel does not build up enough 
stress to allow events with large $N_{f}$ to occur.  Smaller events continually
relieve the accumulated stress instead.
For example, with the small number of flux paths
in Fig.~\ref{fig:3}(b), events with large $N_{f}$ are rare, and $\alpha \sim
3.4$.  Fig.~\ref{fig:3}(a) corresponds to the extreme case of a sample
with only one channel, for which $n_{p}=0.96/\lambda^{2}$ and
$\alpha \sim 4.4$.  In samples with high pin density, even after the stress
in one vortex path has been depleted by a large avalanche, other areas 
still contain enough stress to remain active in large and
small events while the depleted regions build up stress again.  
This leads to a greater likelihood of large events and correspondingly smaller
values of $\alpha$, $\alpha \lesssim 2.4$.  
Distributions similar to this high pin density case
have been obtained experimentally in
\cite{4}, where values of $\alpha$ ranging from $1.4$ to $2.2$
are observed.  The pinning density from grain
boundaries in the experimental sample is very high, 
$n_{p} \sim 100/\lambda^{2}$,
so it is reasonable that the $\alpha$ values observed in \cite{4}
are similar to the $\alpha$ values produced by our most densely 
pinned samples.  Broad distributions with $\alpha$'s
of $1.7$ to $2.2$ were also observed in \cite{7}, 
in good agreement with both \cite{4} and our results.  
In addition, \cite{7} finds a regime where avalanches of a
characteristic size occur, offering an experimental example
in which samples with a lower density of weaker pinning sites
produce narrower distributions, as we also observe.

We have quantitatively shown how pinning 
determines the nature of vortex avalanches. By using
large-scale MD simulations, we image pulses of motion 
in chain-like disturbances through the sample.
The presence or absence of distinct
channels for flow leads to a crossover from broad distributions of avalanche
size to characteristic sizes.  Pinning strength causes a transition
between strongly plastic flow to mildly plastic, ``semi-elastic'' flow.
Lowering the pinning density causes a transition from broad distributions
to distributions with characteristic sizes.

Computer services were provided by: the Maui High Performance Computing
Center, sponsored in part by the Phillips Laboratory, Air Force Materiel
Command, USAF, under cooperative agreement number F29601-93-2-0001;
and by the University of Michigan 
Center for Parallel Computing, partially funded by NSF grant
CDA-92-14296.

\end{document}